\documentclass[conference]{IEEEtran}
\IEEEoverridecommandlockouts
\usepackage{cite}
\usepackage{amsmath,amssymb,amsfonts}
\usepackage[algo2e]{algorithm2e}
\usepackage{graphicx}
\usepackage{textcomp}
\usepackage{xcolor}
\usepackage{amsmath}
\SetKwInput{KwInput}{Input}                % Set the Input
\SetKwInput{KwOutput}{Output} 
\usepackage{algorithm,algorithmic,amsmath}
\def\BibTeX{{\rm B\kern-.05em{\sc i\kern-.025em b}\kern-.08em
    T\kern-.1667em\lower.7ex\hbox{E}\kern-.125emX}}
\usepackage{optidef}
\begin{document}

\title{Joint Matrix Completion and Compressed Sensing for State Estimation in Low-observable Distribution System\\

}

\author{Shweta~Dahale,~\IEEEmembership{Student Member,~IEEE,}
       Balasubramaniam~Natarajan,~\IEEEmembership{Senior~Member,~IEEE}% <-this % stops a space
\thanks{S. Dahale and B. Natarajan are with Electrical and Computer Engineering, Kansas State University, Manhattan, KS-66506, USA, (e-mail:
sddahale@ksu.edu, bala@ksu.edu). This material is based upon work supported by the Department  of  Energy,  Office  of  Energy  Efficiency  and  Renewable Energy  (EERE),  Solar  Energy  Technologies  Office,  under Award Number DE-EE0008767}}

\maketitle

\begin{abstract}
Limited measurement availability at the distribution grid presents challenges for state estimation and situational awareness. This paper combines the advantages of two sparsity-based state estimation approaches (matrix completion and compressive sensing) that have been proposed recently to address the challenge of unobservability. The proposed approach exploits both the low rank structure and a suitable transform domain representation to leverage the correlation structure of the spatio-temporal data matrix  while incorporating  the power-flow constraints of the distribution grid. Simulations are carried out on three phase unbalanced IEEE 37 test system to verify the effectiveness of the proposed approach. The performance results reveal - (1) the superiority over traditional matrix completion and (2) very low state estimation errors for high compression ratios representing very low observability. 
\end{abstract}


\section{Introduction}
Distribution grid operation is becoming more challenging due to an increase in the penetration of distributed energy resources. Instances of reverse power flow and undesired voltage rise has increased and will occur more frequently in the future. Therefore, state estimation (SE) is critical for the monitoring and control of a distribution grid. However, extending the conventional SE approaches for the  distribution grid is difficult mainly because the  system is  highly unobservable at the grid edge. The estimation of the network states with only a limited number of measurements is a major challenge.   Furthermore, the various characteristics of distribution grids such as low $X/R$ ratio \cite{dehghanpour2018survey}, limited  bandwidth capacity, unbalanced operation \cite{hayes2014state}, and cyber-security issues hinders the successful adoption of conventional SE  approaches in distribution grids.
\par
Weighted least squares (WLS) estimation  represents the conventional approach for distribution system state estimation (DSSE). To address the low-observability issue of distribution system, historical data based pseudo-measurements are used along-with the WLS approach. However, the inaccuracies in pseudo-measurements impacts the state estimation performance \cite{clements2011impact}.  Data driven approaches proposed in \cite{pertl2016voltage} %\cite{manitsas2012distribution}
employ model-free structure to estimate the voltage states. However, these approaches still require large  number of PMUs to be installed. Recently,  sparsity-based approaches have been used for DSSE to address the challenge of unobservability. These approaches exploit the network structure to estimate the states at the current levels of measurement availability.  These approaches neither require pseudo-measurements nor any extra metering devices. Compressive sensing (CS) based DSSE was one of the first sparsity-based solution proposed in \cite{shafiul}. This approach exploits the spatial and temporal sparsity of measurements in a linear transformation basis \cite{joshi2018framework}, \cite{karimi2017compressive}.
%An extension of CS based DSSE for three-phase unbalanced system is proposed in \cite{karimi2017compressive}. 
Matrix completion (MC) \cite{donti2019matrix} based DSSE is an another alternative to deal with limited system observability. This method leverages the standard matrix completion along-with the power-flow constraints to acknowledge the physical network constraints.  A comparative analysis of these sparsity based approaches for DSSE along-with their robust formulation is presented in \cite{9247106}. 
Authors in \cite{dahale2020multi} proposes a Gaussian process based approach along-with matrix completion to deal with multi time-scale measurements in a smart distribution system. 
%Another class of sparsity based DSSE is tensor completion which fills the missing elements in a tensor by suitable low-rank approximation \cite{madbhavi2020enhanced}. 
\par
In order to accurately estimate the states with high probability using the sparsity based approaches, the requirement of minimum number of measurements must be satisfied. In compressive sensing, the reconstruction of length $N$ states using $M$ measurements where $M \ll N$ is possible by exploiting the sparsity of states in a transformed basis. $K$-sparse states are recovered accurately using $M \geq c K \mbox{log}(N/K)$ i.i.d gaussian measurements \cite{baraniuk2007compressive}. In a matrix completion approach, the minimum number of measurements ($m$) required to recover the matrix of size $n_1 \times n_2$  with high probability is $m \geq Cn^{1.2} r \mbox{log}n$ where $n = max(n_1, n_2)$ and $r$ is the rank of matrix \cite{candes2009exact}. These requirements restrict the application of sparsity based approaches for highly unobservable distribution system. Furthermore, in the matrix completion approach the minimization of nuclear norm requires solving a semi-definite program which becomes computationally inefficient for large matrices. An efficient alternating minimization algorithm is proposed in \cite{liu2020matrix} that reformulates the matrix completion problem with time-series data. 
%However, this approach only exploits the low-rank property and hence, the recovery accuracy is poor at low fractions of the available data (FAD). 
\par
This paper proposes two unique approaches to estimate the system states when  availability of spatio-temporal measurements at the local control center is very limited. We consider a commonly occurring practical scenario where the sensors at specific spatial locations send data to the local center at a particular sampling rate. The first proposed approach estimates the states by performing matrix completion across space and compressive sensing across time using an alternating minimization approach. This formulation estimates the states in a single shot by exploiting the low rank property as well as temporal sparsity in the measurements while incorporating the power-flow constraints.
\par
In the second approach, the compressive sensing and matrix completion are performed in two stages. In the first stage, the compressed measurements from a single sensor are recovered by exploiting the sparsity of the states in a linear transformation basis. In the second stage, matrix completion across the network is performed at individual time instants. 
The contributions of this paper are as follows:
\begin{itemize}
    \item An efficient and unique state estimation formulation that combines matrix completion and compressive sensing based approaches in a single powerflow constrained optimization framework is proposed for the first time.
    
    \item The proposed approach exploits the low rank property and compactness of temporal data in the wavelet transform domain.  We validate that both these property (low-rank and DCT compactness) hold true for practical data.
    
    \item The proposed algorithm effectively estimates the states with high fidelity in very low observability region. We demonstrate the performance of the algorithm for IEEE 37 unbalanced test system. Relative to the classical matrix completion approaches, the error
     performance of the proposed approach offer nearly 91\% improvement at 10\% of the fraction of available measurements.
\end{itemize}

\section{Background}
\hspace{-0.5cm} Consider a power distribution grid with $|\mathcal{P}|$ three phase non-slack buses. Sensors are deployed throughout the grid but due to communication and other constraints, only a fraction of data is aggregated from these sensors and used for DSSE. This section first reviews the classic matrix completion and compressive sensing approaches for DSSE. 

\subsection{Classic Matrix completion}
In a classic matrix completion based approach for DSSE, a structured matrix $\mathbf{M}$ is formed such that each column represents a phase and each row represents a measurement associated with the phase of each bus. The matrix is given as,
\begin{equation}
       [\Re(v_i),\Im(v_i ),|v_i |,\Re(s_i ),\Im(s_i )]^\intercal
       \label{matrixeqn1}
\end{equation}
where, $s_i$ and $v_i$ represents the apparent power injection and voltages of the $i^{th}$ bus respectively. The term $\Re(\cdot)$ and $\Im(\cdot)$ represents the real and imaginary part of a complex variable respectively.   

In a distribution system, the matrix $\mathbf{M}$ is partially observed. At the local control center, only a subset of the measurement matrix i.e. $P_{\Omega}(\mathbf{M})$ is available. The goal is to fill the missing entries in the matrix by exploiting the relationship among the raw measurements. Specifically, the missing elements of the matrix  are filled by suitable low-rank approximation augmented with the power-flow constraints \cite{9247106}. The corresponding optimization formulation is given as, 
\begin{equation}
\begin{aligned}
\mathbf{\hat{X} }= 
\underset{\mathbf{X} \in \mathbb{R}^{5 \times |\mathcal{P}|} }{\text{\enspace argmin}}
\enspace \| \mathbf{X} \|_* \\
\text{subject to} 
&  & \| P_{\Omega}({\mathbf{M}}) - P_{\Omega}(\mathbf{X}) \|_F^2 < \epsilon\\ 
\end{aligned}
\label{equation9}
\end{equation}
\begin{equation}
    \mathbf{v \approx B} \begin{bmatrix}\Re(\mathbf{s})\\ \Im(\mathbf{s}) \end{bmatrix} + \mathbf{w},
    \label{equ_pf2}
\end{equation}
\begin{equation}
    \mathbf{|v| \approx C\mathbf} \begin{bmatrix}\Re(\mathbf{s})\\ \Im(\mathbf{s}) \end{bmatrix} + \mathbf{|w|},
    \label{eq4}
\end{equation}
where, (\ref{equ_pf2}) and (\ref{eq4}) captures  the linearized power-flow constraints given in \cite{bernstein2017linear}.
%given as,
 %$$
  % \mathbf{B}  = \Big(\mathbf{Y_{LL}^{-1}} \text{diag} (\mathbf{\bar{w}})^{-1},
  % -j\mathbf{Y_{LL}^{-1}} \text{diag} (\mathbf{\bar{w}})^{-1}\Big) \text{ and,} 
%$$$$
 %  \mathbf{C}  = \Big(\mathbf{|\text{diag}(\bar{w})|^{-1} \Re(\text{diag}(\bar{w}) B}\Big) \text{ and,} 
%$$$$
 %   \mathbf{w}  = -\mathbf{Y_{LL}^{-1}} \mathbf{Y_{L0} v_0} \text{\enspace is the zero-load voltage}
%$$
%where, $\mathbf {v_0}$ denote the complex vectors collecting the three phase nodal voltage at the slack bus.
%Here, $\mathbf{Y_{LL}} \in$ $\mathbb{C}^{|\mathcal{P}| \times |\mathcal{P}|}$ and $\mathbf{Y_{L0}} \in \mathbb{C}^{|\mathcal{P}| \times 3}$ are the sub-matrices of the three-phase admittance matrix,\\
%\begin{equation}
%\mathbf{Y} = 
%\begin{bmatrix}
%    \mathbf{Y_{00}}  &  \mathbf{Y_{0L}} \\
 %   \mathbf{Y_{L0}} &  \mathbf{Y_{LL}}
%\end{bmatrix}
%\in \mathbb{C}^{3(|\mathcal{P}|+1) \times 3(|\mathcal{P}|+1)}
%\end{equation}
Here, the nuclear norm $\| \mathbf{X} \|_* = \sum_{i=1}^{r = min(5,|\mathcal{P}|)} \sigma_i (\mathbf{X})$  is the sum of the singular values of the matrix $\mathbf{X}$.
While matrix completion exploits the spatial correlation by low rank approximation, it fails to  capture the temporal correlation of the states. Another sparsity based approach that effectively captures spatial or temporal correlation is compressive sensing discussed in II.B.

\subsection{Classic compressive sensing}
Compressive sensing based DSSE exploits the temporal or spatial sparsity of measurements or states in a linear transformation basis. %Let the states $\mathbf{z} = [\Re(\mathbf{s}),\Im(\mathbf{s}),\Tilde{\mathbf{v}}]^\intercal \in \mathbb{R}^{N}$ be compressible in a linear transformation basis such that,
Let the states $\mathbf{z} \in \mathbb{R}^{N}$ be compressible in a linear transformation basis such that,
\begin{equation}
   \mathbf{z}  = \mathbf{D a} 
   %=  \begin{bmatrix}
  % \Psi_p \enspace \mathbf{0} %\enspace \mathbf{0} \\
  %  \mathbf{0} \enspace \Psi_q \enspace \mathbf{0}\\
   % \mathbf{0} \enspace \mathbf{0} %\enspace \Psi_v 
  % \end{bmatrix} 
   %\mathbf{a}
\end{equation}
where \textbf{a} has at most $K$ $\ll$ $N$ significant coefficients i.e., \textbf{z} 
is $K$-sparse in sparsifying basis $\mathbf{D}$. 
Compressed measurements are achieved by taking $M \ll N$  random projections of $\mathbf{z}$,

\begin{equation}
 \mathbf{h} = \mathbf{\Phi} \mathbf{z};
  \mathbf{h} \in \mathbb{R}^{M}, 
\mathbf{\Phi} \in \mathbb{R}^{M \times N},
\label{equation4}
\end{equation}
where, $\mathbf{\Phi}$ is a random measurement/projection matrix (e.g.,
matrix elements distributed as i.i.d. Gaussian random variable with mean $0$ and variance $1/M$ or Bernoulli random
variables). 

In spatial CS, the states $\mathbf{z} =  [\Re(\mathbf{s}),\Im(\mathbf{s}),\Tilde{\mathbf{v}}]^\intercal$ can be estimated
by solving the following $l_1$ minimization problem

\begin{equation}
\begin{aligned}
\mathbf{\hat{a}} = & \underset{\mathbf{s} }{\text{\enspace min}}
\enspace \| \mathbf{s} \|_1 \\
& \text{subject to}
\enspace \mathbf{\| h  - \Phi D s \|_2^2} < \epsilon \\
\end{aligned}
\label{equation6}
\end{equation}
\begin{equation}
    \mathbf{v \approx B} \begin{bmatrix}\Re(\mathbf{s})\\ \Im(\mathbf{s}) \end{bmatrix} + \mathbf{w},
    \label{equ_pf}
\end{equation}
The recovered states $\mathbf{\hat{z}}$ are given as $ \mathbf{\hat{z}} = \mathbf{D} \mathbf{\hat{a}} $.
Here, $\|\mathbf{s} \|_1 $ represents the $l_1$ norm and $\mathbf{B}$, $\mathbf{w}$ are as defined earlier. The compression of the states $\mathbf{z}$ is indicated by the compressed measurement ratio (CMR) given as $CMR = \frac{M}{N}$.
%\begin{equation}
%    CMR = \frac{M}{N}
 %   \label{cmr_eq}
%\end{equation}
In spatial CS, in order to construct the projection matrix $\mathbf{\Phi}$, the elements of $\mathbf{z}$ should be known apriori. However, it may not be practical to construct this matrix. Another form of CS captures the temporal sparsity of each states $\mathbf{z}$  whose optimization formulation is similar to (\ref{equation6}) except the powerflow constraints (\ref{equ_pf}). Thus, it is makes sense to incorporate matrix completion in space and compressive sensing in time to exploit the sparsity of the states. Jointly incorporating both of these approaches would aid in accurately recovering the states in low-observable conditions.

%The result of the optimization problem in (\ref{equation6}) provides an exact reconstruction with overwhelming probability \cite{candes2008introduction} if there exists a $\delta$ $\in$ (0, 1) such that, 
%\begin{equation}
%  (1-\delta)  \| \mathbf{s} \|^2_2 \leq  \| \mathbf{\Phi \Psi s} \|^2_2 \leq (1+\delta)  \| \mathbf{s} \|^2_2,
%\end{equation}
%holds for all $K$-sparse signal \textbf{s}. This is called the Restricted Isometry property (RIP) of order $K$. 
\section{Proposed Approach}
In this section, we have proposed two approaches to jointly estimate the states using  matrix completion and compressive sensing based DSSE techniques. 
\subsection{Joint MC-CS approach}
%In this section, we propose a new formulation to recover the elements of the partially observed matrix $P_{\Omega}(X)$. 
\begin{figure}[h!]
\centering
\includegraphics[width= 0.4\textwidth]{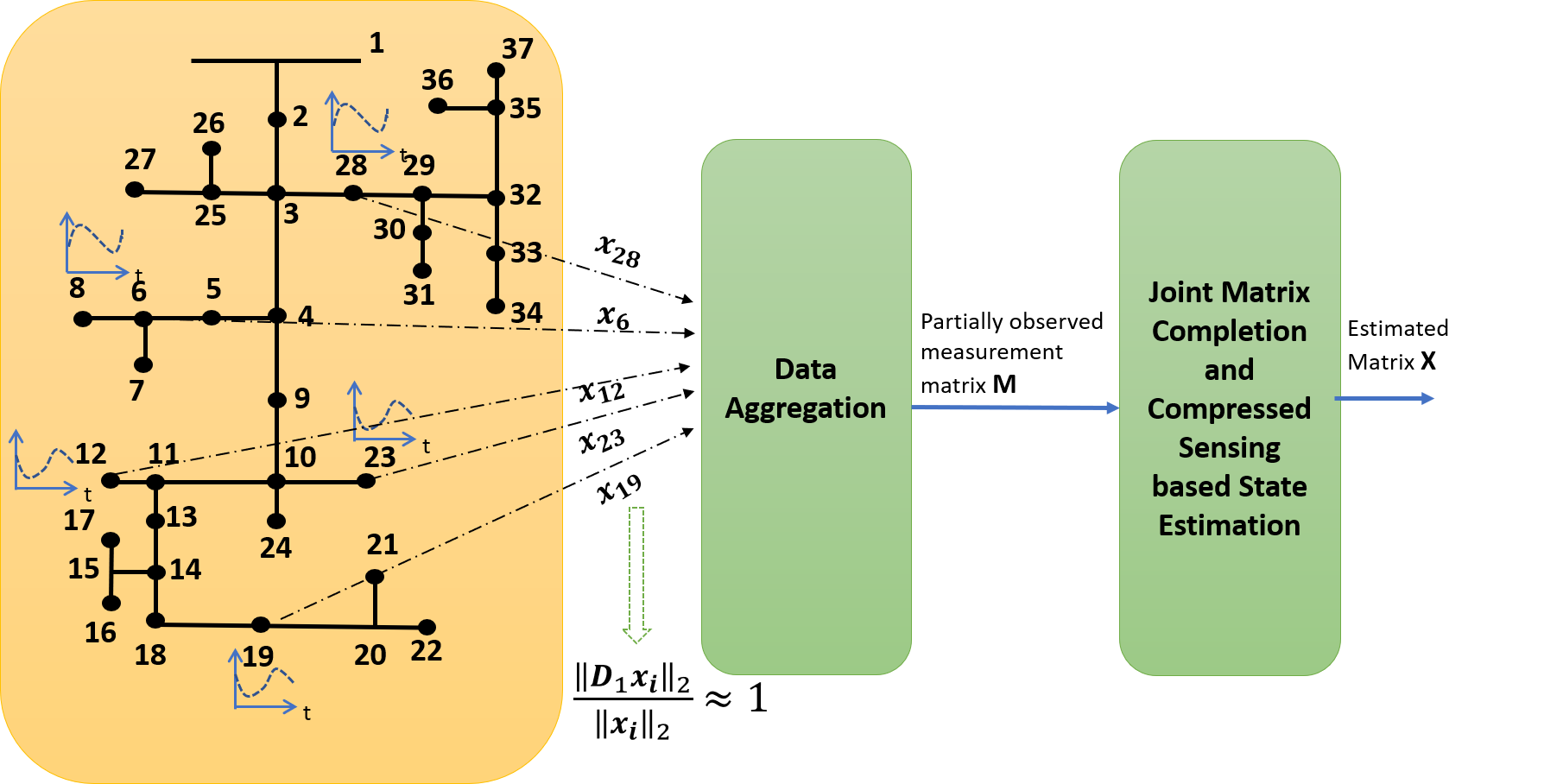}
  \caption{{Framework of the Joint MC-CS approach}}
     \label{fig:my_label23}
 \end{figure}
The framework of the joint MC-CS approach is shown in Fig.\ref{fig:my_label23}.
Assume the sensors located at a subset of buses send data to the local control center at time $t = 1,2,...,T$ at a predefined sampling rate. Let $\mathbf{M}^t$ denote the measurement matrix at time $t$  whose structure is given in (\ref{matrixeqn1}).
A block matrix $\mathbf{M}$ is constructed as, 
\begin{equation}
\mathbf{M} = [\mathbf{M}^1; \mathbf{M} ^2;...,\mathbf{M} ^T] \in \mathbb{R}^{m \times n}
\label{Eq.24}
\end{equation}
Here, $m = 5T$ and $n = |\mathcal{P}|$. %^However,  the columns/rows of the matrix $X$ possess spatial and temporal correlation as well as power-flow constraints. 
%The distribution system is typically unobservable at the grid edge. Hence, only a subset of the measurement matrix $P_{\Omega}(M)$ is available at the grid edge. The recovery of this matrix is possible by exploiting its low rank feature and discrete cosine transform (DCT) compactness property described as, 
The linearized powerflow constraints from (\ref{equ_pf2}) and (\ref{eq4}) at time $t = 1,..,T$
can be written as, 
$$\mathbf{y \approx Ap + b} $$
$$\text{where}: \mathbf{y}= [\Re(\mathbf{v}^1),\Im(\mathbf{v}^1), |\mathbf{v}^1|, ...,\Re(\mathbf{v}^T),\Im(\mathbf{v}^T), |\mathbf{v}^T|]^\intercal,$$
\begin{equation*}
\mathbf{A} = 
\begin{bmatrix}
\mathbf{B_1} & \mathbf{B_2} & \mathbf{0} &\cdots & \mathbf{0} & \mathbf{0} \\
\mathbf{B_3} & \mathbf{B_4} & \mathbf{0} & \cdots & \mathbf{0} & \mathbf{0} \\
\mathbf{C_1} & \mathbf{C_2} & \mathbf{0} & \cdots & \mathbf{0} & \mathbf{0} \\
\vdots  & \vdots  & \vdots & \ddots & \mathbf{0} & \mathbf{0}  \\
\mathbf{0} & \mathbf{0} & \mathbf{0} & \cdots & \mathbf{B_1} & \mathbf{B_2} \\
\mathbf{0} & \mathbf{0} & \mathbf{0} & \cdots & \mathbf{B_3} & \mathbf{B_4} \\
\mathbf{0} & \mathbf{0} & \mathbf{0} & \cdots & \mathbf{C_1} & \mathbf{C_2} \\
\end{bmatrix},
\end{equation*}
\begin{equation*}
\mathbf{p} = [\Re(\mathbf{s}^1)^\intercal \enspace  \Im(\mathbf{s}^1)^\intercal ...\enspace \Re(\mathbf{s}^T)^\intercal \enspace  \Im(\mathbf{s}^T)^\intercal]^\intercal \end{equation*}
\begin{equation*}
 \text{and} \enspace \mathbf{b} = [\Re(\mathbf{w})^\intercal,\Im(\mathbf{w})^\intercal, |\mathbf{w}|^\intercal ... \Re(\mathbf{w})^\intercal,\Im(\mathbf{w})^\intercal, |\mathbf{w}|^\intercal]^\intercal
 \end{equation*}
where, the term $[\mathbf{B_1 \enspace B_2}] = \Re(\mathbf{B})$, $[\mathbf{B_3 \enspace B_4}] = \Im(\mathbf{B})$ and  $[\mathbf{C_1 \enspace C_2}] = \mathbf{C}$.

%However, instead of simply exploiting the low-rank feature of matrix $X^t$, we propose the recovery of matrix $X$ by exploiting the  properties such as, 

It is important to note that the  rows/columns of the matrix $\mathbf{M}$ is observed to exhibit low rank feature and discrete cosine transform (DCT) compactness properties as discussed next.

\subsubsection{ Low rank property} 
    %The missing entries in the incomplete matrix is filled by a suitable low rank approximation. %The matrix $M$ is structured as  such that the columns represents the different buses of the grid and the rows represents the measurement types. 
    The columns of the matrix $\mathbf{M}$ are dependent on each other as there exists spatial correlation between different locations in a power grid. Furthermore, the physics of power-flow relates the different measurements. Hence, $\mathbf{M}$ possess low-rank property which can be evaluated by calculating the SVD of the matrix as, 
    \begin{equation}
      \mathbf{M = U \Sigma V^T}
    \end{equation}
    where the matrix $\mathbf{U}$ $\in$ $\mathbb{R}^{m \times m}$, $\mathbf{V} \in \mathbb{R}^{n \times n}$ and $\mathbf{\Sigma} \in \mathbb{R}^{m \times n}$ containing $p = min(m,n)$ singular values arranged in descending order ($\sigma_1 > \sigma_2 > ... > \sigma_p$).
    For our experiments involving practical data from IEEE 37 bus test system, it can be inferred that the largest 5 singular values occupy about 99.9\% of the energy confirming the low-rank property of the measurement matrix. This property has also been confirmed by other prior efforts on matrix completion \cite{donti2019matrix}. 
    
    %as shown in Fig.\ref{fig:my_label1}. The plot between the weights of the singular values ($g(r)$) vs the rank $r$ is shown in Fig.\ref{fig:my_label1} where the term $g(r)$ is given as,
   % $$g(r) = \frac{\sum_{i=1}^{r} \sigma_i}{\sum_{i=1}^{p} \sigma_i},$$
    %where, $1 \leq r \leq p$. 
\subsubsection{DCT compactness analysis}     
 In a distribution network, the loads are observed to be slowly changing over time. The temporal data in the matrix $\mathbf{M}$ represented by $\mathbf{x}_i$ is observed to exhibit sparsity in a linear transformation basis \cite{shafiul}. Discrete cosine transform (DCT) enables to represent the data in a fewer coefficients. The DCT matrix $\mathbf{D}$ = '$\{d(k,n)\}$' of dimension $T \times T$ as defined in \cite{jain1989fundamentals} can be split as, 
  \begin{equation*} \mathbf{D}=\begin{bmatrix}\mathbf{D}_{1}\\ \mathbf{D}_{2} \end{bmatrix}  \end{equation*}
where $\mathbf{D}_1$ consists of first $'j'$ rows of $\mathbf{D}$ and $\mathbf{D}_2$ consists of last $'T-j'$ rows. To exhibit temporal sparsity for the timeseries data $\mathbf{x}_i$, only few DCT coefficients will capture most of the energy i.e.,
%\begin{equation}
%\frac{\| \mathbf{D}_{1} \mathbf{x}_{i} \|_{2}}{\| \mathbf{x}_{i}\|_{2}} \approx 1
%\end{equation}
%\begin{equation}
%\frac{\| \mathbf{D}_{2} \mathbf{x}_{i} \|_{2}}{\| \mathbf{x}_{i}\|_{2}} \approx 0
%\end{equation}
 %The DCT matrix $\mathbf{D}$ = '$\{d(k,n)\}$' of dimension $T \times T$ as defined in \cite{jain1989fundamentals} is given as, 
%  \begin{align}
%  d(k,n) =
 % \begin{cases}
 % \frac{1}{\sqrt{T}} , \enspace k=0, \enspace 0\leq n \leq T-1
 % \\
 % \frac{2}{\sqrt{T}} cos\big(\frac{\pi(2n+1)k}{2T}\big), \enspace  1\leq k \leq T-1,
 % \\ \hspace{2.9cm} 0\leq n \leq T-1
  %\end{cases}
%\end{align}
\begin{equation*}
\frac{\| \mathbf{D}_{1} \mathbf{x}_{i} \|_{2}}{\| \mathbf{x}_{i}\|_{2}} \approx 1, \enspace \frac{\| \mathbf{D}_{2} \mathbf{x}_{i} \|_{2}}{\| \mathbf{x}_{i}\|_{2}} \approx 0 \end{equation*}
This property can be observed from the practical time-series data from IEEE 37 bus test system where 1-2 DCT coefficients occupy 99\% of the energy, thus proving the temporal data in matrix $\mathbf{M}$ is compact. However, due to limited system observability, only limited entries of the matrix $\mathbf{M}$ are observed. In order to recover the complete matrix, the low-rank feature property, DCT compactness and the linearized power-flow constraints are exploited in an integrated optimization formulation corresponding to,
\begin{equation}
\begin{aligned}
\underset{\mathbf{X}} {\text{\enspace min}}
\enspace \| \mathbf{X}\|_* + \lambda_1 \| P_\Omega (\mathbf{X}) - P_\Omega(\mathbf{M})\|_F ^2 + \nu \|\mathbf{ y - (Ap + b)} \|_2^2 +  \\
\lambda_2 \| \mathbf{s} \|_2 \\
\textrm{s.t.} \quad
\mathbf{y} = [a_1^\intercal \mathbf{X} \enspace a_2^\intercal \mathbf{X}  \enspace ...\enspace a_{3T-2}^\intercal \mathbf{X} \enspace a_{3T-1}^\intercal \mathbf{X} \enspace a_{3T}^\intercal \mathbf{X} ]^\intercal,
\\
\mathbf{p} = [c_1^\intercal \mathbf{X} \enspace c_2^\intercal \mathbf{X} \enspace ...\enspace c_{2T-1}^\intercal \mathbf{X} \enspace c_{2T}^\intercal \mathbf{X}]^\intercal,
\\
\mathbf{s} = 
\begin{bmatrix}
\mathbf{D}_2 \enspace  (e_4^\intercal \enspace reshape (\mathbf{X}(:,1),[5,T]))^\intercal \\

\mathbf{D}_2 \enspace (e_5^\intercal \enspace reshape (\mathbf{X}(:,1),[5,T]) )^\intercal \\

\mathbf{D}_2 \enspace  (e_1^\intercal \enspace reshape (\mathbf{X}(:,1),[5,T]))^\intercal \\

\vdots \\

\mathbf{D}_2 \enspace  (e_1^\intercal \enspace reshape (\mathbf{X}(:,n),[5,T]))^\intercal \\
\end{bmatrix}
\end{aligned}
\label{equation11}
\end{equation}
where, 
$m= 5T$, $n = |\mathcal{P}|$, $\mathbf{X} \in \mathbb{R}^{m \times n}$,  $\mathbf{y} \in \mathbb{R}^{\frac{3}{5}mn}$, $\mathbf{p} \in \mathbb{R}^{\frac{2}{5}mn}$ \\
Here, $a_{3(t-1) + i }= e_{5(t-1)+i}$ and  $c_{2(t-1) + i }= e_{5(t-1)+3+i} $ are the standard basis vectors in $\mathbb{R}^m, $
$e_1$, $e_4$ and $e_5$ are the standard basis vectors in $\mathbb{R}^5$. The parameters $\lambda_1 \geq 0 $, $\nu \geq 0$, $\lambda_2 \geq 0$ are the tuning parameters. 
\\
The matrix $\mathbf{X}$ can be factorized into two matrices $\mathbf{U}$ and $\mathbf{V}$. The nuclear norm of $\mathbf{X}$ can be expressed by the Frobenius norm of matrix $\mathbf{U}$ and $\mathbf{V}$ given as,
\begin{equation}
\begin{aligned}
{\| \mathbf{X}\|_*} = & \underset{\mathbf{U},\mathbf{V} }{\text{\enspace min}} \enspace
 \| \mathbf{U}\|^2_F + \| \mathbf{V} \|^2_F \\
 & \text{subject to} \enspace \mathbf{X = UV}
\end{aligned}
\label{eq14}
\end{equation}
Substituting (\ref{eq14}) in (\ref{equation11}), we obtain the following optimization problem,  
\begin{equation}
\begin{aligned}
\underset{\mathbf{U,V}}
{\text{min}} \enspace
\|\mathbf{U}\|^2_F +\| \mathbf{V} \|^2_F + \lambda_1 \| P_\Omega(\mathbf{UV}) - P_\Omega(\mathbf{ M)}\|_F ^2 + \\
\nu \| f_1(\mathbf{UV}) - (A f_2(\mathbf{UV}) + \mathbf{b}) \|_2^2 +  \lambda_2 \| f_3(\mathbf{UV})\|_2 \\
\textrm{s.t.} \quad
f_1(\mathbf{UV}) = [a_1^\intercal \mathbf{X} \enspace a_2^\intercal \mathbf{X}  \enspace ...\enspace a_{3T-2}^\intercal \mathbf{X} \enspace a_{3T-1}^\intercal \mathbf{X} \enspace a_{3T}^\intercal \mathbf{X} ]^\intercal,
\\
f_2(\mathbf{UV}) = [c_1^\intercal \mathbf{X} \enspace c_2^\intercal \mathbf{X} \enspace ...\enspace c_{2T-1}^\intercal \mathbf{X} \enspace c_{2T}^\intercal \mathbf{X}]^\intercal,
\\
f_3(\mathbf{UV}) = 
\begin{bmatrix}
\mathbf{D}_2 \enspace  (e_4^\intercal \enspace reshape (\mathbf{X}(:,1),[5,T]))^\intercal \\

\mathbf{D}_2 \enspace (e_5^\intercal \enspace reshape (\mathbf{X}(:,1),[5,T]) )^\intercal \\

\mathbf{D}_2 \enspace  (e_1^\intercal \enspace reshape (\mathbf{X}(:,1),[5,T]))^\intercal \\

\vdots \\

\mathbf{D}_2 \enspace  (e_1^\intercal \enspace reshape (\mathbf{X}(:,n),[5,T]))^\intercal \\
\end{bmatrix}
\\ 
\mathbf{X= UV}
\end{aligned}
\label{equation15}
\end{equation}
The matrix completion formulation in (\ref{equation15}) is a  non-convex problem. But using alternating minimization algorithm \cite{liu2020matrix}, the problem becomes convex when either $\mathbf{U}$ or $\mathbf{V}$ is fixed. This algorithm updates the variables $\mathbf{U}$ and $\mathbf{V}$ at each iteration $k$ in an alternating fashion while fixing the other factor. The update rules are given by the following update equations as,
\begin{mini}
  [3]
  {\mathbf{U}^{(k)}}
  {\| \mathbf{U}\|^2_F + \lambda_1 \| P_\Omega (\mathbf{UV}^{(k-1)}) - P_\Omega(\mathbf{M})\|_F ^2 + \nu}  {}{} 
  \breakObjective{\| f_1(\mathbf{UV}^{(k-1)}) - (\mathbf{A} f_2(\mathbf{UV}^{(k-1)}) + \mathbf{b}) \|_2^2} {}{}
  \breakObjective{+ \lambda_2 \| f_3(\mathbf{UV}^{(k-1)})\|_2}{}{}
  {}
  {}
  \label{eqn18}
\end{mini}
\begin{mini}
  [2]
  {\mathbf{V}^{(k)}}
  {\| \mathbf{V}\|^2_F + \lambda_1 \| P_\Omega (\mathbf{U}^{k}\mathbf{V})- P_\Omega(\mathbf{M})\|_F ^2 + \nu}  {}{} 
  \breakObjective{\| f_1(\mathbf{U}^{k}\mathbf{V}) - (\mathbf{A} f_2(\mathbf{U}^{k}\mathbf{V}) + \mathbf{b}) \|_2^2} {}{}
  \breakObjective{+ \lambda_2 \| f_3(\mathbf{U}^{k}\mathbf{V})\|_2 }{}{}
  {}
  {}
   \label{eqn19}
\end{mini}
The alternating minimization approach for the proposed approach is given in Algorithm 1.

\subsection{CS-MC Approach}
The joint MC-CS approach proposed in section III.A directly incorporates the raw information from each of the sensors at a particular bus for state estimation. However, due to the network bandwidth limitation, it may not be practical to collect all the temporal measurements. Furthermore, as stated earlier, to solve (\ref{Eq.24}) involving large matrices can be computationally inefficient.   Therefore, we propose a state estimation approach that alleviates these drawbacks. \par
\begin{figure}[h!]
\centering
\includegraphics[width= 0.4\textwidth]{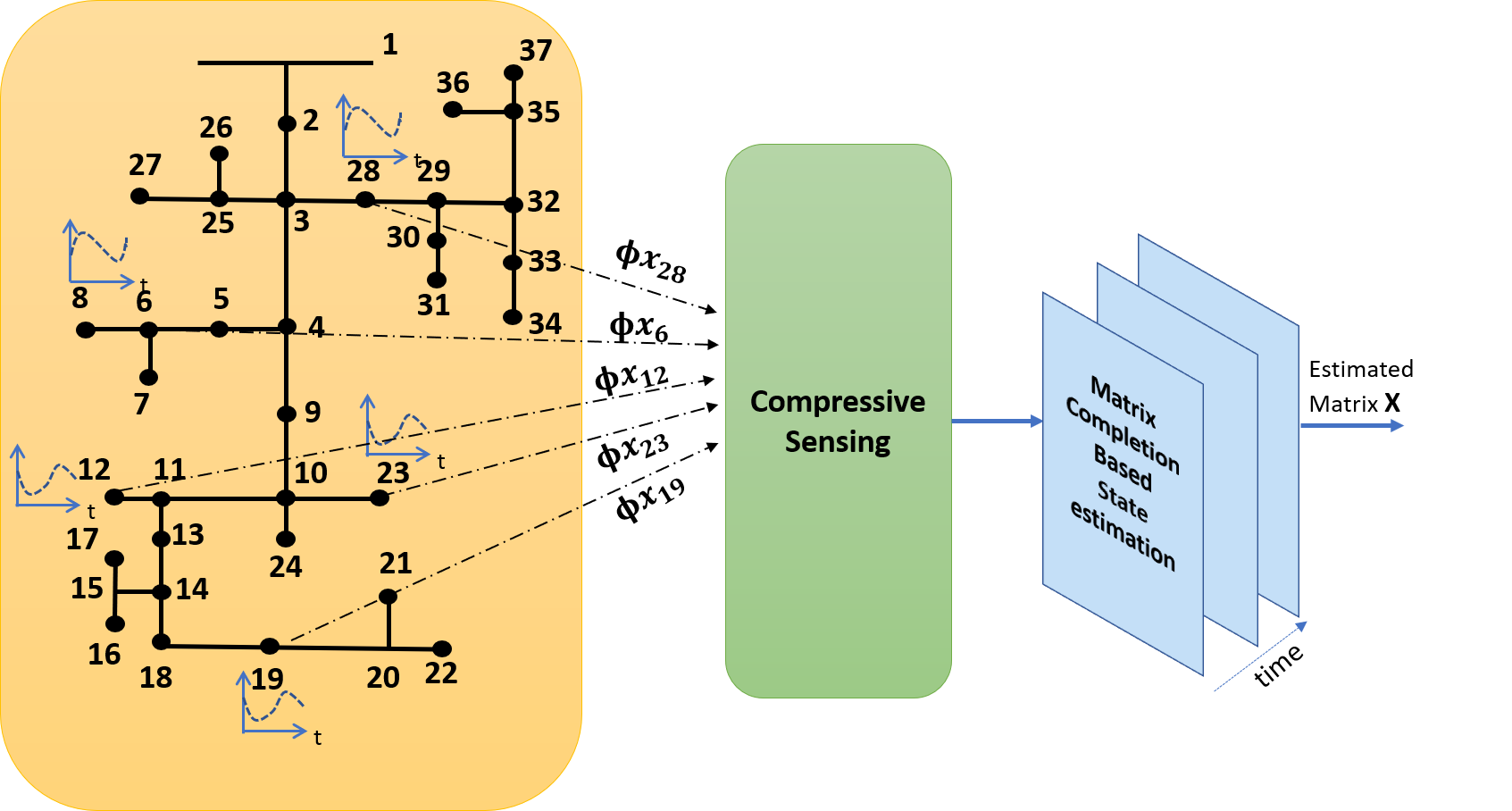}
   \caption{{Framework of the CS-MC approach}}
    \label{fig:my_label18}
\end{figure}

The framework of the CS-MC approach is shown in Fig.\ref{fig:my_label18}.  Assume the measurements collected from each buses  is $\mathbf{\Phi} \mathbf{x}$ where  $\mathbf{\Phi} \mathbf{x} \ll$ $\mathbf{x}$. The recovery of the states is performed in two stages. Firstly, the incomplete temporal measurements are recovered by compressive sensing using (\ref{equation6}). It should be noted that no power-flow constraints are used at this stage.  Once the estimates of all the temporal measurements are obtained, the second stage consists of recovering the spatial states  by classic matrix completion based state estimation using (\ref{equation9}). The state estimation is performed separately at each time steps along-with the power-flow constraints in (\ref{equ_pf2}) and (\ref{eq4}). 
%Algorithm 2 summarizes the  CS-MC approach.

\begin{algorithm}
\KwInput{Measurement matrix $\mathbf{M}$, Set of known indices $\Omega$, DCT matrix $\mathbf{D}$, system model $\mathbf{A}$, $\mathbf{b}$, number of iterations $N$.}
 initialization:\ Compute the SVD of the matrix $ \mathbf{ M = U \Sigma V^T}$, set $\mathbf{V}^{(0)} = \mathbf{\Sigma}^{0.5}\mathbf{V}$
  \begin{algorithmic}[1]
    \FOR{$k = 1,..,N$}
\STATE Solve (\ref{eqn18})

\STATE Solve (\ref{eqn19})
    \ENDFOR
    \STATE \textbf{return} $\mathbf{X} = \mathbf{U}^{(N)}\mathbf{V}^{(N)}$
  \end{algorithmic}
  \caption{Alternating minimization Algorithm for matrix completion across space and compressive sensing across time}
\end{algorithm}
%\begin{algorithm}
%\KwInput{Compressed measurements $\mathbf{h}$, Set of known indices $\Omega$, DCT matrix $\mathbf{D}$, system model $\mathbf{B}$, $\mathbf{C}$,   $\mathbf{w}$, random projection matrix $\mathbf{\Phi}$}
% initialization:\ Compute the SVD of the matrix $ \mathbf{ M = U \Sigma V^T}$, set $\mathbf{V}^{(0)} = \mathbf{\Sigma}^{0.5}\mathbf{V}$
 % \begin{algorithmic}[1]
%\STATE Decompress the measurements $\mathbf{h}$ using CS (\ref{equation6}).
%\STATE Form a structured matrix $\mathbf{M}$ for each time instant separately using the decompressed measurements (\ref{matrixeqn1}).
%\STATE Estimate the states of the matrix individually at each time instances by matrix completion based DSSE (\ref{equation9}), (\ref{equ_pf2}), (\ref{eq4}). 
 %   \STATE \textbf{return} $\mathbf{X}$ 
%  \end{algorithmic}
 % \caption{CS-MC Approach}
%\end{algorithm}
\section{Computational Complexity}
In this section, the computational complexity of the proposed approaches for a matrix $\mathbf{X} \in \mathcal{R}^{5T \times |\mathcal{P}|}$ is discussed.  
In the joint MC-CS approach, in order to solve the power-flow constraints, $\mathcal{O}(|\mathcal{P}|^3 T^3)$ computations are required at each iteration. 
In a classic MC approach, the matrix $\mathbf{X}  \in \mathcal{R}^{5 \times |\mathcal{P}|}$ is estimated separately for each time $T$. Denoting $m$ and $n$ as the rows and columns of $\mathbf{X}$, the main computation is involved in calculation of the SVD of the matrix which is  $\mathcal{O}(mn \cdot \mbox{min}({m,n}))$ along-with the power-flow constraints $\mathcal{O}(n^3)$ at each time. Hence, the overall worst case computational complexity is  $\mathcal{O}(|\mathcal{P}|^3 T)$. The CS-MC approach performs CS for each temporal state and classic matrix completion in the next step. CS requires $\mathcal{O}(T \cdot k \cdot \mbox{min}(T,k))$ computations where  $T$ denotes the time and $k$ as constraints. The overall worst case complexity in both the steps is $\mathcal{O}(|\mathcal{P}|^3 T)$. This illustrates that the proposed joint MC-CS approach is computationally expensive as compared to the conventional matrix completion as well as the CS-MC approach. 
%However, the  performance  gains achieved in the joint MC-CS approach is significantly higher than the other two approaches. 
\section{Simulation Results and Discussion}
\begin{figure}[h!]
\centering
\includegraphics[width= 0.35\textwidth]{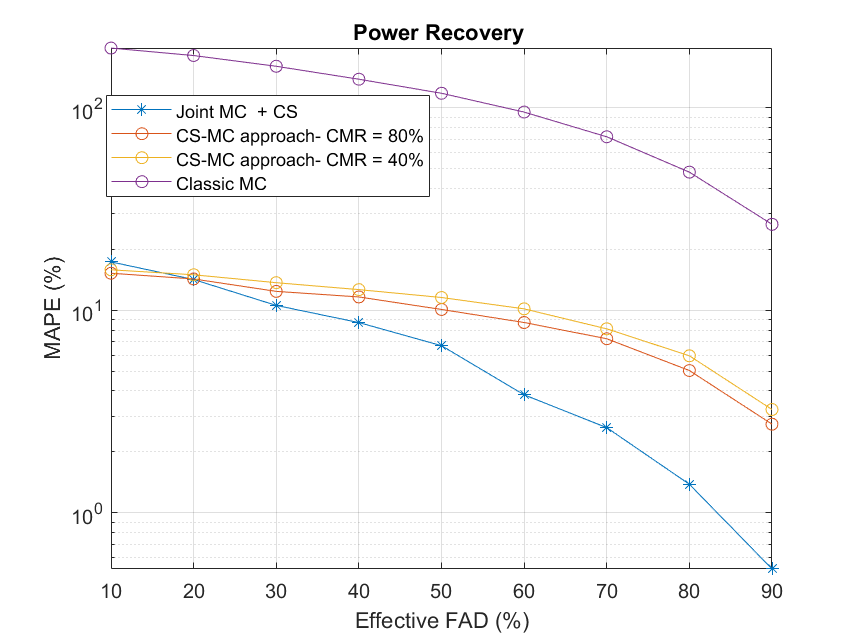}
   \caption{{Power recovery at different FADs}}
     \label{fig:my_label3}
 \end{figure}

\begin{figure}[h!]
\centering
\includegraphics[width= 0.35\textwidth]{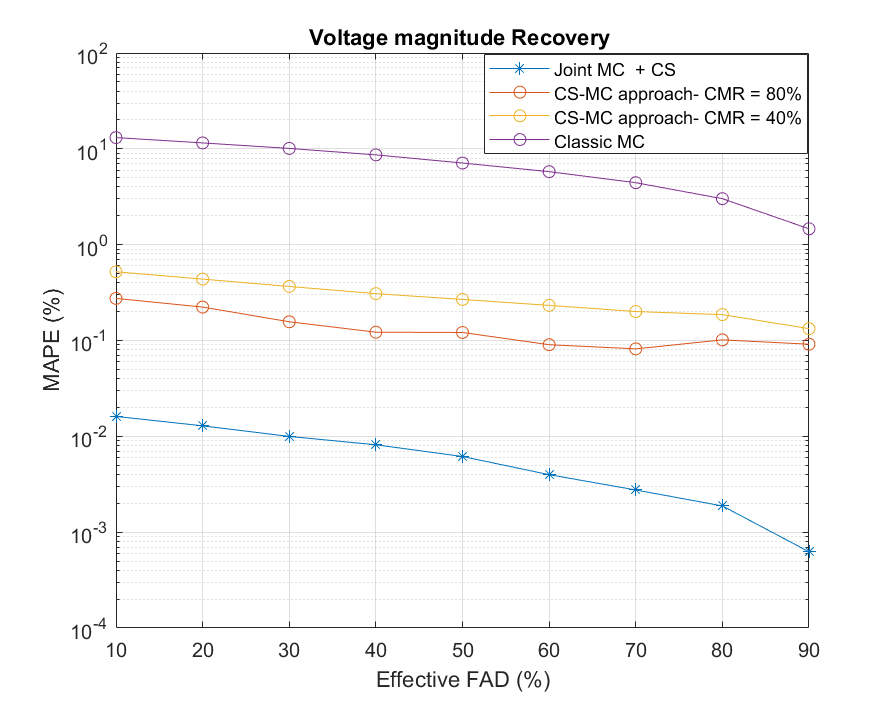}
   \caption{{Voltage magnitude recovery at different FADs}}
     \label{fig:my_label4}
 \end{figure}

\begin{figure}[h!]
\centering
\includegraphics[width= 0.35\textwidth]{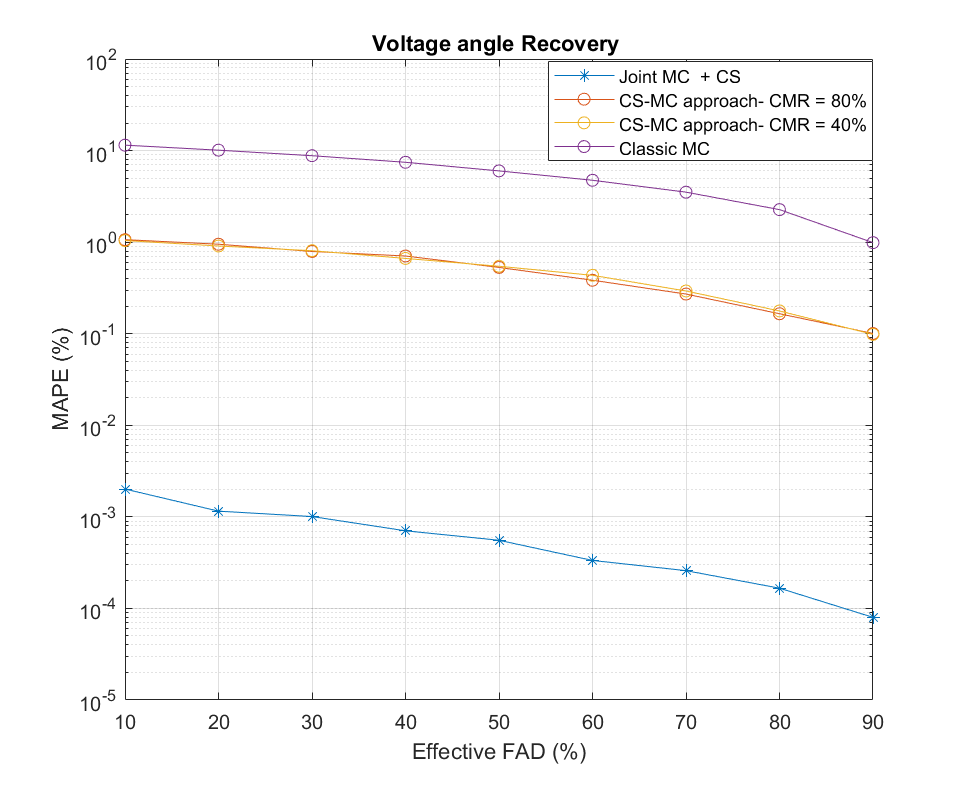}
   \caption{{Voltage angle recovery at different FADs}}
     \label{fig:my_label5}
 \end{figure}

In this section, the efficacy of the proposed formulation is demonstrated on the IEEE 37 unbalanced three phase test system. We characterize the performance of power and voltage magnitude recovery using the mean absolute percentage error (MAPE) metric and the voltage angle recovery using mean integrated absolute error (MIAE) metric \cite{9247106}.
For simulation, we consider a matrix that includes 8 time steps and whose entries are randomly available representing different fractions of available data (FAD). The compression of the temporal measurements in the CS-MC approach is indicated by CMR as defined earlier. 
Fig. \ref{fig:my_label3} shows the recovery performance of power (active and reactive) using the two proposed approaches. 
Fig.\ref{fig:my_label4} -\ref{fig:my_label5} shows the comparative performance of the proposed formulation with the classic MC in the recovery of voltage magnitude and voltage angle states respectively. It can be inferred that the proposed joint MC-CS approach as well as the CS-MC approach outperforms the classic matrix completion at all FADs. The performance is superior especially in the low observability region. This is due to the fact that this approach exploits both the spatial correlation as well as temporal correlation of the states. The joint MC-CS approach, although computationally expensive, is superior than 
CS-MC approach. This is due to the fact that raw measurements are used for estimating the states rather than the compressed measurements. The recovery of states in CS-MC approach also depends on the CMR of the temporal measurements. The error in the recovery performance increases as the CMR is decreased. Fig. \ref{fig:my_label3}-\ref{fig:my_label5} shows the performance of CS-MC at 40\% and 80\% CMR. 
\vspace{-0.5cm}
\section{Conclusion}
This paper presents an efficient joint matrix completion and compressive sensing approach to perform DSSE. The proposed approach employs an alternating minimization approach to estimate an incomplete spatio-temporal matrix. The efficacy of the proposed approach was demonstrated using the IEEE 37 bus test system. It can be inferred from the simulation results that the proposed approach has a superior performance relative to classic matrix completion and provides lower state estimation errors at high compression ratios.  
%\section{Acknowledgement}
%This material is based upon work supported by the Department  of  Energy,  Office  of  Energy  Efficiency  and  Renewable Energy  (EERE),  Solar  Energy  Technologies  Office,  under Award Number DE-EE0008767.
\bibliographystyle{IEEEtran}
\bibliography{ref.bib}
\end{document}